\documentclass[aps,prx,twocolumn,superscriptaddress,showpacs]{revtex4}
\usepackage{graphicx}
\usepackage{latexsym}
\usepackage{amssymb}
\usepackage{amsmath}
\usepackage{amsfonts}
\usepackage{bm}
\usepackage{multirow}
\usepackage{color}
\usepackage{xcolor}

\usepackage[colorlinks=true, citecolor={blue!80!black}, urlcolor={blue!50!black}, linkcolor = {blue!80!black}]{hyperref}

\usepackage[percent]{overpic}
\usepackage{tabularx}

\usepackage{comment}

\newcommand{\SU}{\mathrm{SU}}
\newcommand{\U}{\mathrm{U}}
\newcommand{\Sp}{\mathrm{Sp}}

\newcommand{\beq}{\begin{equation}}
\newcommand{\eeq}{\end{equation}}
\newcommand{\beqn}{\begin{eqnarray}}
\newcommand{\eeqn}{\end{eqnarray}}

\DeclareMathAlphabet{\mathbbold}{U}{bbold}{m}{n}

\def\SU{{\rm SU}}

\def\U{{\rm U}}

\definecolor{forestgreen}{rgb}{0.13, 0.55, 0.13}

\newcommand{\mcS}{\mathcal{S}}

\begin{document}

\title{Emergent Fermi surface in a triangular-lattice SU(4) quantum antiferromagnet}

\author{Anna Keselman}
\affiliation{Kavli Institute for Theoretical Physics, University of California, Santa Barbara, CA 93106-4030}

\author{Bela Bauer}
\affiliation{Microsoft Station Q, Santa Barbara, California 93106-6105, USA}

\author{Cenke Xu}
\affiliation{Department of Physics, University of California,
Santa Barbara, CA 93106, USA}  

\author{Chao-Ming Jian}
\affiliation{Microsoft Station Q, Santa Barbara, California 93106-6105, USA}

\begin{abstract}

Motivated by multiple possible physical realizations, we study the
$\SU(4)$ quantum antiferromagnet with a fundamental representation
on each site of the triangular lattice. We provide evidence for a
gapless liquid ground state of this system with an emergent
Fermi surface of fractionalized fermionic partons coupled with a
$\U(1)$ gauge field. Our conclusions are based on numerical
simulations using the density matrix renormalization group (DMRG)
method, which we support with a field theory analysis.

\end{abstract}

\maketitle

Realizations of quantum spin liquids --- quantum phases of spins
whose ground state
is not described by local ordering patterns but instead
characterized by exotic quantum entanglement --- have been highly
sought-after since such phase was first hypothesized~\cite{QSLreviews}.
Within the broad family of spin liquids, a particularly elusive
category are gapless spin liquids that exhibit gapless excitations
on an extended region in the momentum space, akin to the Fermi
surface in ordinary metals.
The known realizations of such gapless phases in systems of $\SU(2)$
spins usually require complicated Hamiltonians beyond the
Heisenberg interaction, such as ring exchange
terms~\cite{Motrunich2007,Sheng2008,Sheng2009,block2011,Block2011b,Mishmash2011,Jiang2013,He2018},
staggered chiral three-spin
interactions~\cite{pereira2017gapless,bauer2019}, or antiferromagnetic Kitaev
interactions in an external field~\cite{hickey2019emergence,patel2019magnetic,jiang2019field}.

Here, we report strong evidence for a gapless liquid with an
emergent Fermi surface of fractionalized partons in the
nearest-neighbor SU(4) Heisenberg quantum antiferromagnet on the
triangular lattice with a fundamental representation on each site.
While SU($N$) antiferromagnets were suspected to harbor exotic
phases already in the early days of the
field~\cite{readsachdev,SLsachdev,sachdevread1,sachdevread2,sachdevread3,rokhsar,Penc2003}
and recent work has demonstrated the presence of a Dirac spin
liquid in the same model on the honeycomb
lattice~\cite{Corboz2012}, our motivation for studying this model
stems primarily from the availability of several possible
experimental realizations. In transition metal oxides, spin and
orbital degrees of freedom may be described by an effective
$\SU(4)$ quantum magnet~\cite{SU41,SU42,SU43}. Cold atomic gases
formed by atoms with large hyperfine spin component can form
effective $\SU(N)$ quantum antiferromagnet~\cite{xusun}, and
spin-3/2 atoms can naturally form $\Sp(4)$ or $\SU(4)$ quantum
antiferromanget~\cite{wu1,wu2,wu3} when only the $s$-wave
scattering between the atoms is considered. Most recently, it was
also proposed that some of the $2d$ systems with Moir\'{e}
superlattices may be described by an approximate $\SU(4)$ quantum
antiferromagnet~\cite{xuleon,senthil,senthil2,xuFM,fumag} at
commensurate fillings where correlated insulators were observed
recently~\cite{wangmoire,mag01,young2018}.

In the following, we will first introduce a parton mean-field construction for a candidate liquid state for the model. We then
carefully examine the properties of this state when placed on
quasi-one-dimensional cylinder geometries, including the effects
of symmetry-allowed perturbations specific to these geometries.
These will also be the target of unbiased numerical simulations
using the density-matrix renormalization group (DMRG)
method~\cite{white1992,schollwoeck2005}. We find our numerical
results to be in agreement with predictions from the field theory
that describes the proposed liquid state. For two cases of even
circumference, we find gapped states with ordering patterns which
are consistent with the one-dimensional field theory that contains
relevant symmetry-allowed perturbations deviating from a gapless
fixed point; while in a case with odd circumference, where there
are no relevant translation-symmetric operators, we find a gapless
state whose structure factor exhibits sharp features consistent
with the field theory. We thus conclude that our proposed theory
describes the system accurately in quasi-one-dimensional
geometries and thus likely also in the two-dimensional limit.

\emph{Model-} We study the Kugel-Khomskii model~\cite{Kugel1982}
on the two-dimensional triangular lattice at the SU(4) symmetric
point
\begin{equation}
H = J \sum_{\langle i j \rangle} \left( 2 {\bf S}_i \cdot {\bf
S}_j + \frac{1}{2} \right) \left( 2 {\bf V}_i \cdot {\bf V}_j +
\frac{1}{2} \right),
    \label{eq:HeisenbergModel}
\end{equation}
where $J>0$ is an antiferromagnetic coupling, and ${\bf S}_i$
(${\bf V}_i$) denote the $S=1/2$ spin (orbital) degrees of freedom
at site $i$. We denote the three Pauli matrices that act on the
two-fold spin (orbital) indices as $\sigma^a$ ($\tau^a$), such
that $S^a=\sigma^a/2$ ($V^a=\tau^a/2$) with $a=x,y,z$. We can view
the degrees of freedom on each site as a pseudospin in the
fundamental representation of SU(4), with the 15 operators $\{
\sigma^a, \tau^b, \sigma^a\tau^b \}_{a,b=x,y,z}$ being the 15
generators of SU(4). The Hamiltonian Eq.~\eqref{eq:HeisenbergModel}
can be interpreted as an SU(4) antiferromagnetic Heisenberg model.

\begin{figure}
    \begin{overpic}[width=0.49\columnwidth]{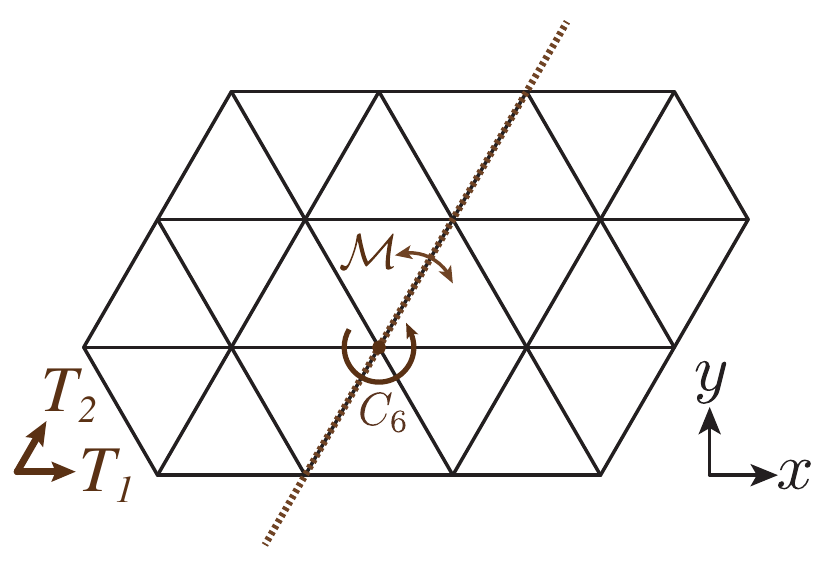} \put (0,60) {\footnotesize{(a)}} \end{overpic}
    \begin{overpic}[width=0.49\columnwidth]{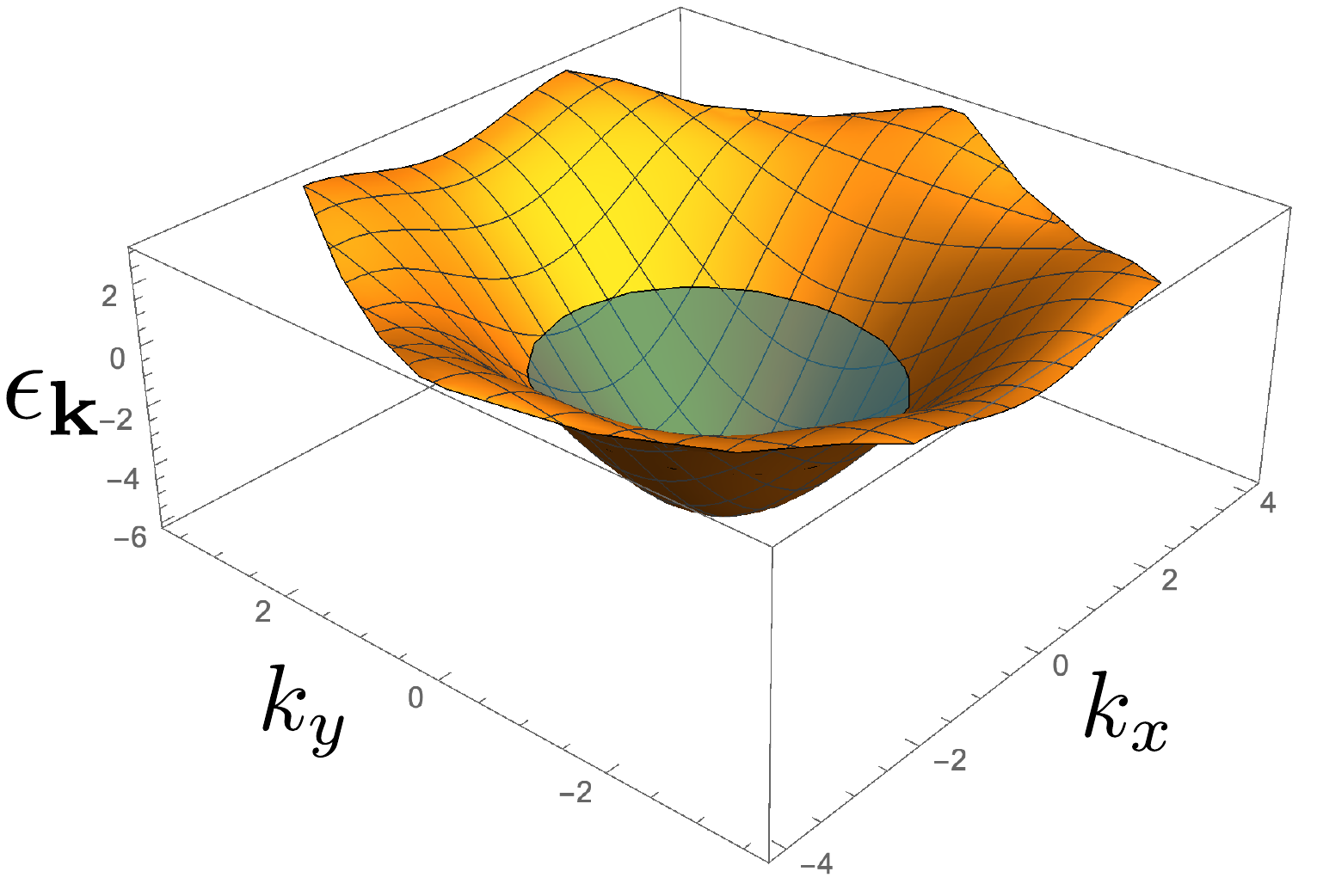} \put (0,60) {\footnotesize{(b)}} \end{overpic}
    \caption{  (a) $T_1$ and $T_2$ denote the translation symmetries along the two basis vectors of the 2d triangular lattice. $\mathcal{M}$ denotes the mirror symmetry with the $T_2$ direction as the mirror plane and $C_6$ denotes the 6-fold crystal rotation symmetry. (b) The parton mean-field band structure (orange), i.e. the single parton energy $\epsilon_{\bf k}$ as a function of crystal momentum $k_{x,y}$, is shown. The Fermi level corresponding to filling $\nu=1/4$ is depicted in blue.   }
    \label{fig:TriangularLattice_Band}
\end{figure}

The Hamiltonian Eq.~\eqref{eq:HeisenbergModel} is invariant under
the global SU(4) pseudospin rotation symmetry, as well as the
spatial symmetries of the triangular lattice including the
translation symmetries $T_{1,2}$, the mirror symmetry
$\mathcal{M}$ and the 6-fold rotation symmetry $C_6$ as shown in
Fig.~\ref{fig:TriangularLattice_Band}(a). In addition, as a
spin-orbital system, the model naturally admits a time-reversal
(TR) symmetry $\mathcal{T}$ whose action depends on the physical
nature of the orbital degrees of freedom. If the orbital space is the valley space in the Moir\'{e} systems where the valleys are exchanged under the TR symmetry, $\mathcal{T}$ acts on the SU(4) pseudospin degrees of degrees of freedom as the operator $i\sigma^y \tau^x \mathcal{K}$ with $\mathcal{K}$ representing the complex conjugation. In systems where the orbitals transform trivially under $\mathcal{T}$, the TR action is given by the operator $i\sigma^y \mathcal{K}$ instead. As we will see, all our discussions below apply to both realizations of the TR symmetry.

\emph{Fermionic parton mean-field ansatz-} We now construct a
candidate for the ground state of the model in Eq.
~\eqref{eq:HeisenbergModel}. We start by introducing a 4-component
fermionic parton on each site, and use $f_{i,m=1,..,4}$ (and
$f^\dag_{i,m}$) to denote the corresponding annihilation (and
creation) operators. The four components of the fermionic parton
can be also labeled by the two-fold spin indices and two-fold
orbital indices. They transform into each other under the global
SU(4) pseudospin rotation. The SU(4) pseudospin operators (on the
site $i$) can be represented in terms of the fermionic parton as
\begin{equation}
    S^a_i=\frac{1}{2}f^\dag_i \sigma^a  f_i,\ V^b_i=\frac{1}{2}f^\dag_i \tau^b  f_i,\  (S^aV^b)_i=\frac{1}{4}f^\dag_i \sigma^a\tau^b  f_i.
    \label{eq:parton_decomposition}
\end{equation}
The physical Hilbert space of SU(4) pseudospins is obtained from
the Hilbert space of the fermionic partons by imposing the
constraint $n_i=\sum_{m=1}^4 f^\dag_{i,m} f_{i,m} = 1$ on each
site $i$.

We consider the simplest parton mean-field ansatz given by the
following mean-field Hamiltonian:
\begin{align}
    H_{\rm mf} = -t \sum_{\langle ij \rangle} \sum_{m=1}^4 f^\dag_{i,m} f_{j,m} + h.c. .
\label{eq:parton_meanfield}
\end{align}
which only contains nearest-neighbor parton hoppings with a
uniform $t>0$ on the triangular lattice. This mean-field ansatz
preserves the full SU(4) pseudospin rotation symmetry, the
space-group symmetries of the triangular lattice, and the
time-reversal symmetry $\mathcal{T}$ that transforms the partons as $f_i \rightarrow i\sigma^y \tau^x f_i$ when the orbitals are physically realized by valleys and as $f_i \rightarrow i\sigma^y f_i$ when the orbitals transform trivially under $\mathcal{T}$.

This mean-field ansatz yields a single 4-fold-degenerate parton
band. At the mean-field level, the singe-occupancy constraint
$n_i=1$, requires the partons have filling factor $\nu = 1/4$ and
hence, results in a parton Fermi surface as shown in
Fig.~\ref{fig:TriangularLattice_Band}(b). Beyond mean-field, the
constraint above can be implemented by a dynamical U(1) gauge
field coupled to the fermionic partons.

\begin{figure}
    \centering
    \begin{overpic}[width=0.49\columnwidth]{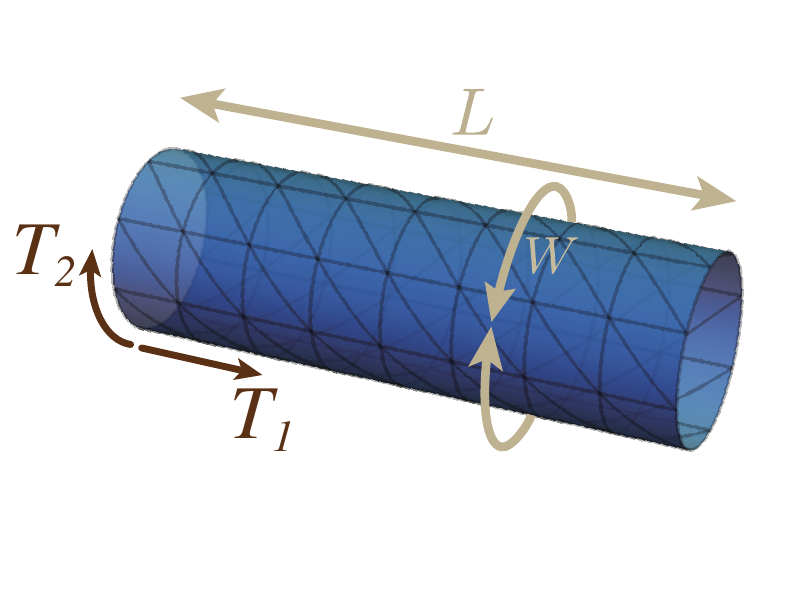} \put (0,60) {\footnotesize{(a)}} \end{overpic}
    \begin{overpic}[width=0.49\columnwidth]{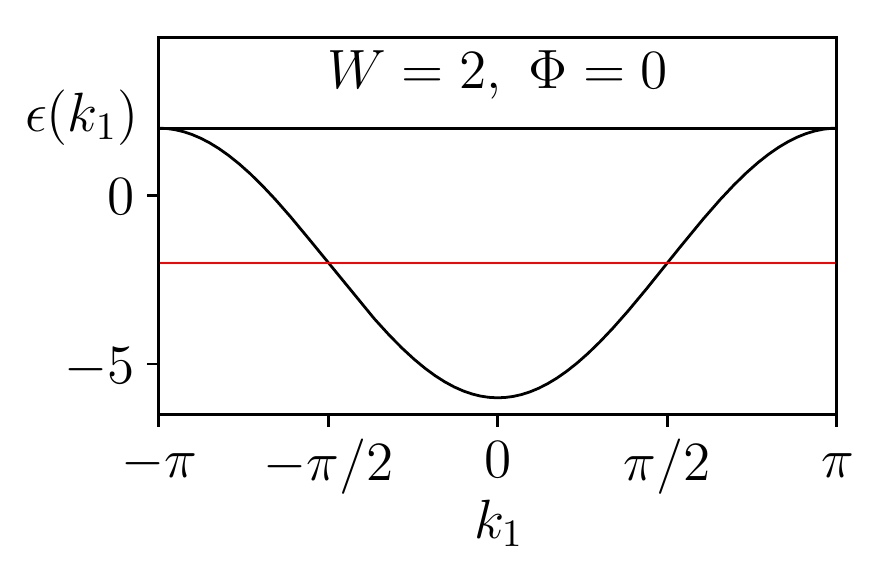} \put (0,60) {\footnotesize{(b)}} \end{overpic} \\
    \begin{overpic}[width=0.49\columnwidth]{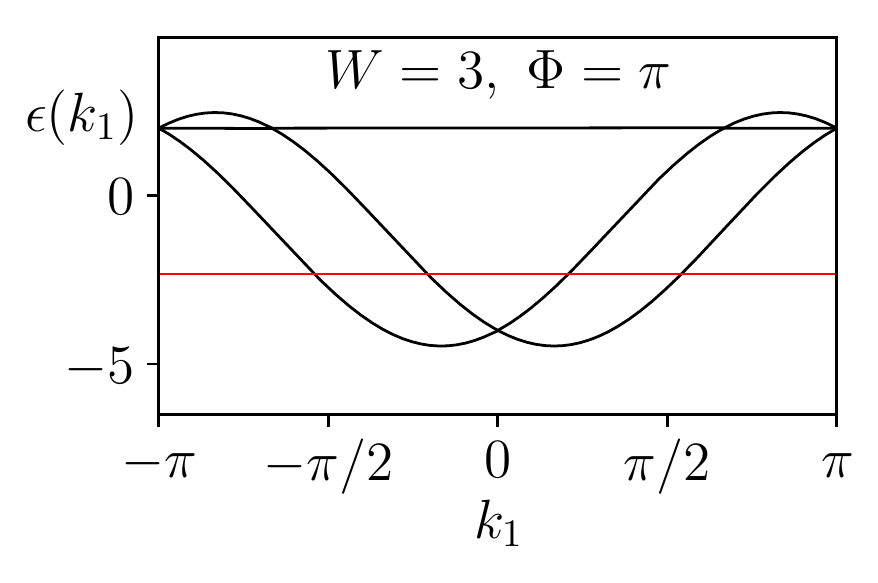} \put (0,60) {\footnotesize{(c)}} \end{overpic}
    \begin{overpic}[width=0.49\columnwidth]{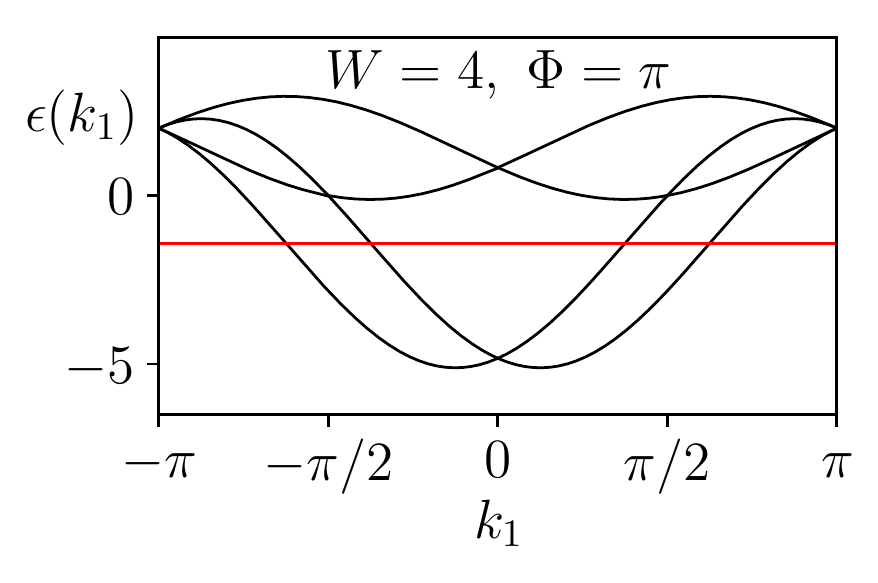} \put (0,60) {\footnotesize{(d)}} \end{overpic}
\caption{(a) Compactification of the 2d lattice along the $T_2$
direction, resulting in a cylinder geometry. (b,c,d) The parton
mean-field band structure (with energies given in the unit of $t$)
on the compactified, quasi-1d geometry, when the number of unit
cells along $T_2$ is $W$=2,3,4, respectively. In this geometry an
additional degree of freedom, the flux $\Phi$ through the cylinder
has to be considered. We plot the band structure for a flux of
$\Phi=0$ for $W=2$, and $\Phi=\pi$ for $W=3,4$.}
    \label{fig:1DGeometry_Band}
\end{figure}

\emph{Finite circumference cylinders-}
Our numerical simulations
will be performed for cylinder geometries that are constructed by
compactifying the $T_2$ direction and imposing periodic boundary
conditions on the SU(4) pseudospin variables (see
Fig.~\ref{fig:1DGeometry_Band}(a)). The circumference of the
cylinder is denoted as $W$ and the length (along the $T_1$
direction) of the cylinder as $L$. The quasi-1d system with finite
$W$ (and infinite $L$) maintains the space-group symmetry
$T_{1,2}$ and $\mathcal{M}$ but breaks the $C_6$ symmetry to a
two-fold crystal rotation symmetry $C_2$.

We can place the mean-field
Hamiltonian~\eqref{eq:parton_meanfield} on the same geometry if we
additionally specify the boundary condition for the partons in the
$T_2$ direction. The only choices that preserve either one of the
TR symmetry or the product of mirror and rotation $\mathcal{M}C_2$
are periodic and antiperiodic boundary conditions. These can also
be interpreted as placing a U(1) gauge flux $\Phi=0$ and
$\Phi=\pi$, respectively, through the cylinder. In general, there
isn't a simple reasoning which value of $\Phi$ is more favorable
for a certain geometry. We can view it as a discrete parameter
(our only parameter) when comparing the parton ansatz and the
results of the DMRG study.

For finite $W$, the two-dimensional parton band structure reduces
to $W$ (4-fold degenerate) one-dimensional bands, each
parameterized by the crystal momentum $k_1$ along the $T_1$
direction. Different one-dimensional bands can be distinguished by
their crystal momentum $k_2$ along the $T_2$ direction.
The parton Fermi level is still determined by the parton filling
constraint $\nu=1/4$. In general, the number of (partially)
occupied one-dimensional parton bands depends on both $W$ and
$\Phi$. In the following, we will focus on the $\Phi=0$ scenario
for $W=2$ and $\Phi=\pi$ for $W=3,4$, as we find that these
choices are most consistent with the DMRG results. The corresponding
one-dimensional band structures are shown in
Fig.~\ref{fig:1DGeometry_Band}(b-d). A more comprehensive
comparison with different choices of $\Phi$ for $W=2,3,4$ is given
in the Supplementary Material~\cite{SM}. The Fermi momenta for
each $W$ can be calculated directly from the mean-field
Hamiltonian Eq.~\eqref{eq:parton_meanfield}. For $W=2$ with
$\Phi=0$, the single partially occupied band has $k_2=0$ and the
$k_1$-values of the Fermi momenta are $\pm \pi/2$. For $W=3,4$
with $\Phi=\pi$, the two bands that are (partially) occupied by
the partons have crystal momenta $k_2 = \pm \pi/W$ and the
$k_1$-values of the four Fermi momenta are $\pm \pi/(2W) \pm \pi
W/8 $. In fact, in all the cases we consider, these Fermi momenta
are also completely fixed by the symmetries, as we show in detail
in the Supplementary Material~\cite{SM}. Pairwise differences of
the Fermi momenta will play an important role in the later
discussion.

For each $W$, by linearizing the parton band structure around each
Fermi point, we can write down a continuum Lagrangian of
low-energy partons in these quasi-1d geometries:
\begin{align}
    \mathcal{L}^{(0)}_{W} = \sum_{r,n,m}  & \left[ \psi^\dag_{r,n,m}\big(i\partial_0 + v_r i\partial_1\big)\psi_{r,n,m} \right].
    \label{eq:L0}
\end{align}
Here $\mu=0,1$ label the temporal and spatial components. The
fermionic fields $\psi_{r,n,m}$ describe the low-energy partons
near the Fermi points, where $m$ is the SU(4) pseudospin index,
$n$ is the band index, and $r=R(L)$ stands for right(left) movers
with a velocity $v_{r,n}=\pm v_n$ respectively. In all the scenarios we consider, the Fermi points in a given geometry are all related by
symmetries ($\mathcal{T}$, $\mathcal{M}$ and $C_2$), so
are the respective velocities. Thus, we find that the Lagrangian
in Eq.~\eqref{eq:L0} describes SU(4)-invariant massless Dirac
fermions for $W=2$, whereas for $W=3,4$ it describes massless
Dirac fermions with an enhanced SU(8) symmetry.

Going beyond the mean-field level, the parton filling constraint,
$n_i=1$, leads to the coupling of the low-energy fermions in
Eq.~\eqref{eq:L0} to a dynamical U(1) gauge field $a_{\mu}$, via
the substitution $i\partial_{\mu} \rightarrow
i\partial_{\mu}-a_{\mu}$. Thus, the low-energy theory for $W=2$
($W=3,4$) is given by the $N_f=4$ ($N_f=8$) QED$_2$, or
equivalently the 1+1d SU(4)$_1$ (SU(8)$_1$) conformal field theory
(CFT), whose energy spectrum is gapless. The Dirac mass terms are
forbidden in all of these cases due to the translation symmetry
$T_1$.

We next consider symmetry-allowed relevant perturbations to these
gapless theories. More specifically, we will focus on possible
Umklapp scatterings for each $W$. Although these perturbations are
not expected to appear in the 2d limit, we will see that they can
change the low-energy physics dramatically for the cases with
finite circumferences we study numerically.

For $W=2$, the distance between the two Fermi points allows for
the following symmetry-preserving Umklapp interaction
\begin{align}
    \mathcal{L}^{\rm int}_{W =2, \Phi=0} =  \left(\sum_{m=1}^4 \psi^\dag_{L,m} \psi_{R,m}  \right)^2 + h.c.,
    \label{eq:2leg_Interaction}
\end{align}
where we suppressed the band index in the fields $\psi^\dag_{L,m}$
and $\psi_{R,m}$ because there is only one (partially) occupied
band. This interaction commutes with $T_1$ because the Fermi
momenta dictate that under $T_1$, $\psi_{L,m} \rightarrow
e^{-i\pi/2} \psi_{L,m},~ \psi_{R,m} \rightarrow e^{i\pi/2}
\psi_{R,m}$. Using the Fierz identity, this Umklapp interaction
can be written as a back-scattering between left-moving and
right-moving primary fields in the SU(4)$_1$ CFT, both carrying
the 6-dimensional representation of SU(4) (see Supplementary
Material~\cite{SM} for more details). In the SU(4)$_1$ CFT, each of such
primary fields has scaling dimension $1/2$. Therefore, the Umklapp
interaction has a scaling dimension $1$ and, hence, is a relevant
perturbation. It can lead to a phase with a finite vacuum
expectation value $\langle \sum_{m=1}^4 \psi^\dag_{L,m} \psi_{R,m}
\rangle $ that gaps out all low-energy degrees of freedom and
spontaneously breaks the $T_1$-translation symmetry by doubling
the unit cell in the $T_1$ direction. Other symmetries stay intact
in this gapped phase. In fact, for $W=2$, doubling of the unit
cell along the $T_1$ direction in a gapped phase is expected due
to the 1d Lieb-Schultz-Mattis constraint for SU(4) spin
chains~\cite{AffleckLieb1986}.

For $W=3$, due to the (relative) positions of the Fermi points,
the symmetry-allowed Umklapp terms are of high orders (i.e. at
least 16) in terms of the low-energy fermion fields. Therefore,
the effect of Umklapp terms here can be neglected, and the
SU(8)$_1$ CFT (or equivantly the $N_f=8$ QED$_2$) remains a good
description of the system. With $W=3$, each unit cell in the $T_1$
direction has three SU(4) pseudospins. In the absence of $T_1$
symmetry breaking, the system has to be gapless based on the SU(4)
Lieb-Schultz-Mattis constraint\cite{AffleckLieb1986}.

For $W=4$, the (relative) positions of the Fermi points allow for
the following symmetry-preserving Umklapp interactions
\begin{align}
\left(\sum_{m} \psi^\dag_{L,n,m} \psi_{R,n,m}  \right) \left(\sum_{m} \psi^\dag_{L,n',m} \psi_{R,n',m}  \right)+h.c.,
    \label{eq:Ny4_Umklapp_Terms}
\end{align}
where $n,n'=1,2$ label the two 1d parton bands that are
(partially) occupied. Again these interactions preserve the
translation $T_1$, as its action is given by $T_1: \sum_{m}
\psi^\dag_{L,n,m} \psi_{R,n,m}\rightarrow -\sum_{m}
\psi^\dag_{L,n,m} \psi_{R,n,m}$ for $n=1,2$. These Umklapp
interactions can all be written as the back-scattering between
left-moving and right-moving primary fields in the SU(8)$_1$ CFT
which both carry the 28-dimensional representation of SU(8) (see
Supplementary Material~\cite{SM} for more details). In the SU(8)$_1$ CFT,
each of such primary fields has a scaling dimension $3/4$.
Therefore, each of these Umklapp interactions has a scaling
dimension $3/2$ and again is a relevant perturbation. These
perturbations can lead to a phase with nonzero expectation value
of $\langle \sum_{m=1}^4 \psi^\dag_{L,n,m} \psi_{R,n,m} \rangle $
(for both $n=1,2$) that gap out the system while breaking the
$T_1$-translation symmetry by doubling the unit cell. Other
symmetries remain intact in this gapped phase. Interestingly, for
$W=4$, each unit cell along the $T_1$ has four SU(4) pseudospins.
Thus, in this case, the SU(4) Lieb-Schultz-Mattis constraint does not require a gapped phase to break the $T_1$ translation symmetry.
As we will demonstrate, the DMRG with $W = 4$ also shows a
doubling of the unit cell, which is consistent with our field theory
analysis.

\emph{Numerical study-} We perform DMRG simulations using the
ITensor library~\cite{ITensor}; to accelerate the simulations, we
explicitly conserve three U$(1)$ quantum numbers corresponding to
total $S^z$, $V^z$, and $S^z V^z$. A key observable is the
pseudospin gap $\Delta$, which we obtain as the energy difference
between the ground states in the sectors with $\left(S^z, V^z, S^z
V^z\right)=\left(0,0,0\right)$ (which contains the SU(4) singlet)
and $\left(S^z, V^z, S^z V^z\right)=\left(1,0,-1/2\right)$. For
each cylinder circumference $W$, we obtain the gap $\Delta$ for
cylinders of varying length and then perform an extrapolation to
the thermodynamic limit.

\begin{figure}
    \begin{tabular}{m{0.49\columnwidth} m{0.49\columnwidth}}
        \begin{overpic}[width=0.49\columnwidth]{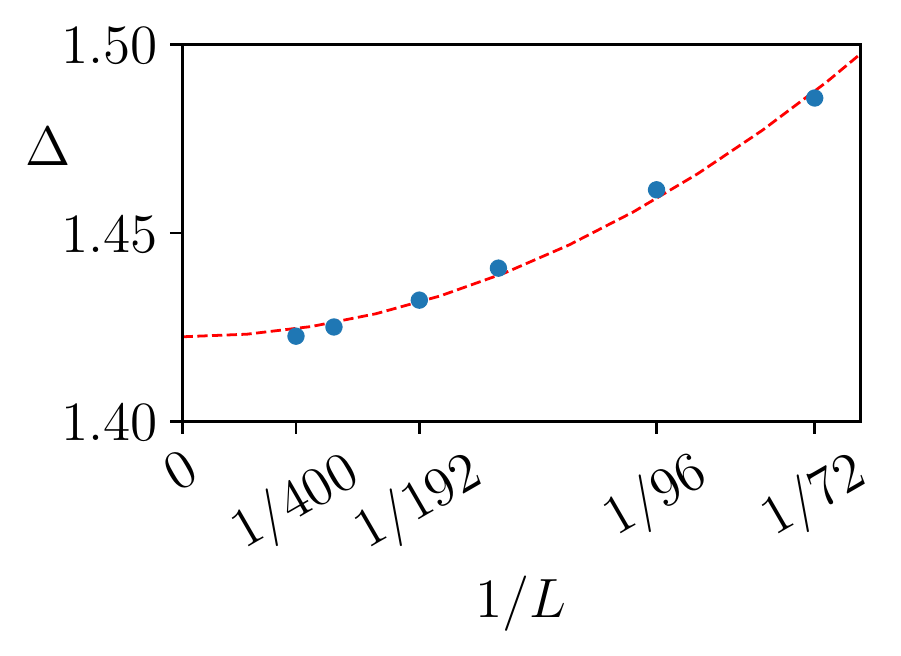} \put (-2,70) {\footnotesize{(a)}} \end{overpic} &
        \begin{overpic}[width=0.49\columnwidth]{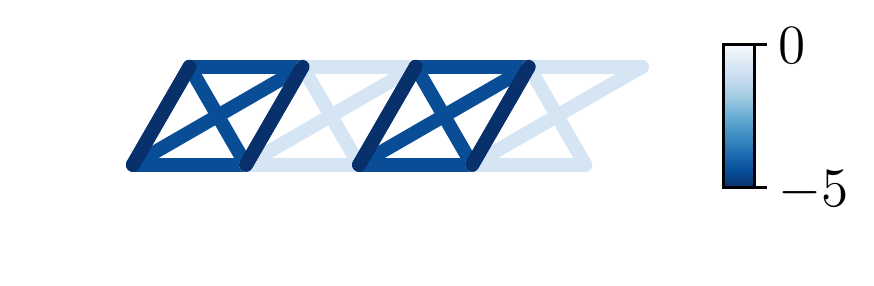} \put (0,43) {\footnotesize{(b)}} \end{overpic} \vspace{5pt} \\
        \begin{overpic}[width=0.49\columnwidth]{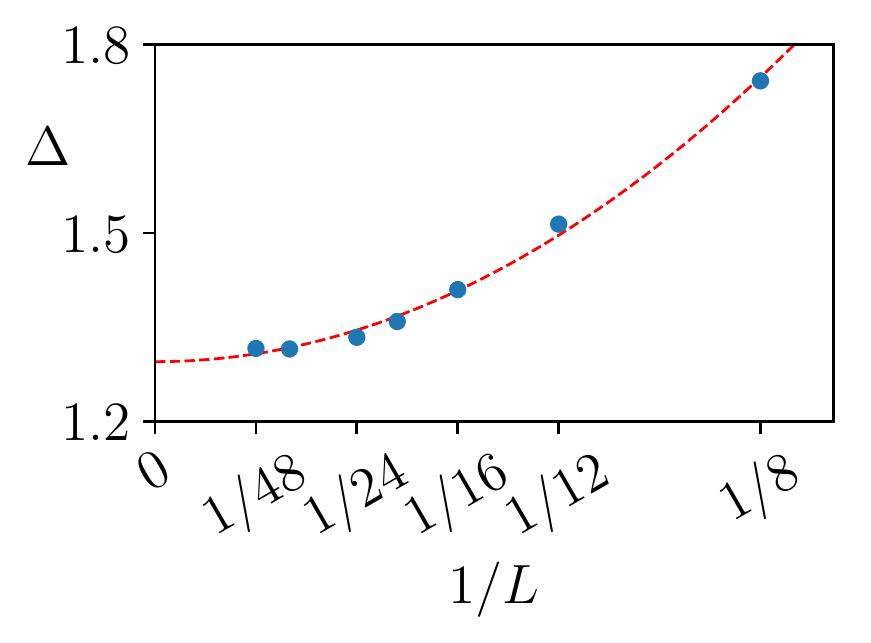} \put (-2,70) {\footnotesize{(c)}} \end{overpic} &
        \begin{overpic}[width=0.49\columnwidth]{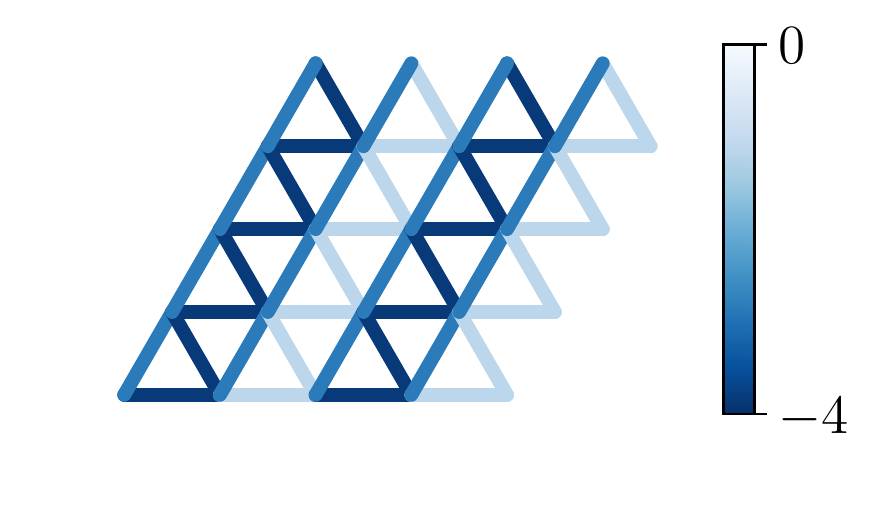} \put (0,57) {\footnotesize{(d)}} \end{overpic}
        \vspace{3pt}
    \end{tabular}
\caption{(a,c) Pseudospin gap as function of inverse system size
for finite cylinders of width $W=2,4$ obtained using a bond
dimension of up to $M=4000$, resulting in truncation errors of
$\epsilon_{\rm tr}\simeq10^{-5}$ ($\epsilon_{\rm
tr}\simeq10^{-9}$) for $W=4$ ($W=2$). Red dashed line in each of
the plots is a fit to $\Delta_0+a/L^2$ yielding $\Delta_0=1.42$
for $W=2$ and $\Delta_0=1.29$ for $W=4$. (b,d) The bond
expectation values for the middle four rungs in a cylinder of
length $L=24$ and width $W=2,4$ respectively.}
    \label{fig:NyEven}
\end{figure}

The gap obtained for $W=2,4$ is shown in
Fig.~\ref{fig:NyEven}(a,c). In both cases, we find that the gap
remains finite in the limit of $L\to\infty$, consistent with the
expectation of a gapped phase due to the Umklapp scattering.
Translation-symmetry breaking can be observed directly in the bond
expectation values $\langle \sum_\alpha \mcS^\alpha_i\cdot
\mcS^\alpha_j \rangle$, where $\mcS^\alpha$ are the 15 SU(4)
pseudospin operators $\{\sigma^a, \tau^b,
\sigma^a\tau^b\}_{a,b=x,y,z}$, and $i,j$ are a pair of
nearest-neighbor sites. The pattern of bond expectation values is
shown for the middle four rungs in a cylinder of length $L=24$ and
circumference $W=2,4$ in Fig.~\ref{fig:NyEven}(b,d). In both cases
one can clearly see that the translation symmetry is broken and
there is a unit cell doubling along $T_1$, in agreement with the
symmetry-breaking pattern expected from the field theory analysis
in the previous section. We emphasize that for $W=4$, no
translation symmetry breaking along the circumference of the
cylinder (i.e., along $T_2$) is observed (see Supplementary
Material~\cite{SM} for more details), indicating that the state does not
originate from plaquette coverings of the lattice as proposed in
Refs.~\onlinecite{vdBossche2001,Penc2003}.

\begin{figure}
    \centering
    \begin{overpic}[width=0.6\columnwidth]{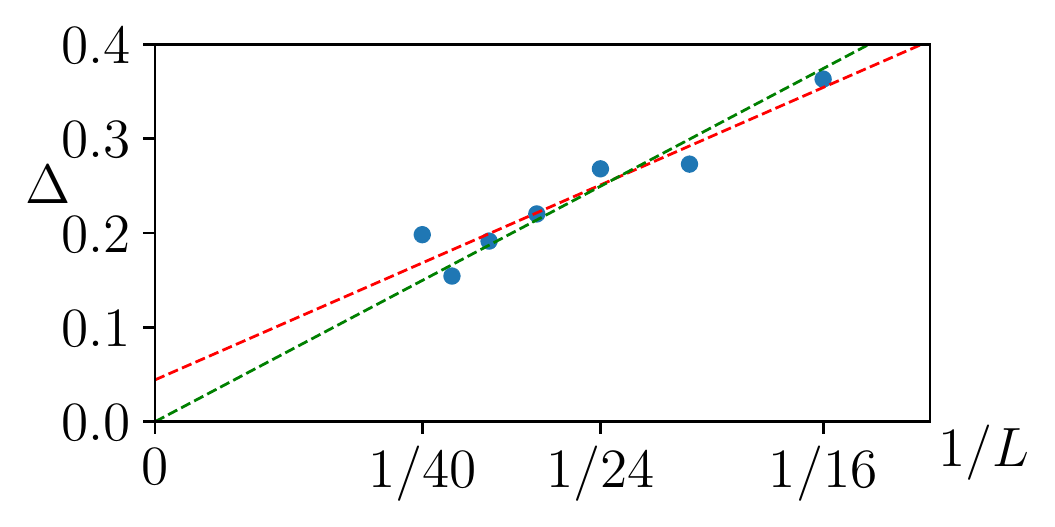} \put (-5,45) {\footnotesize{(a)}} \end{overpic} \\
    \begin{overpic}[width=0.49\columnwidth]{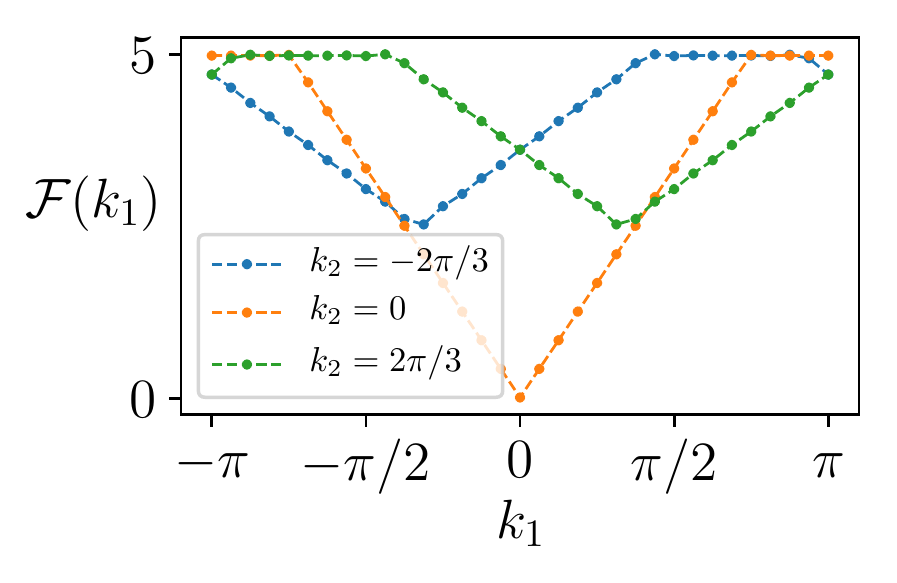} \put (0,62) {\footnotesize{(b)}} \end{overpic}
    \begin{overpic}[width=0.49\columnwidth]{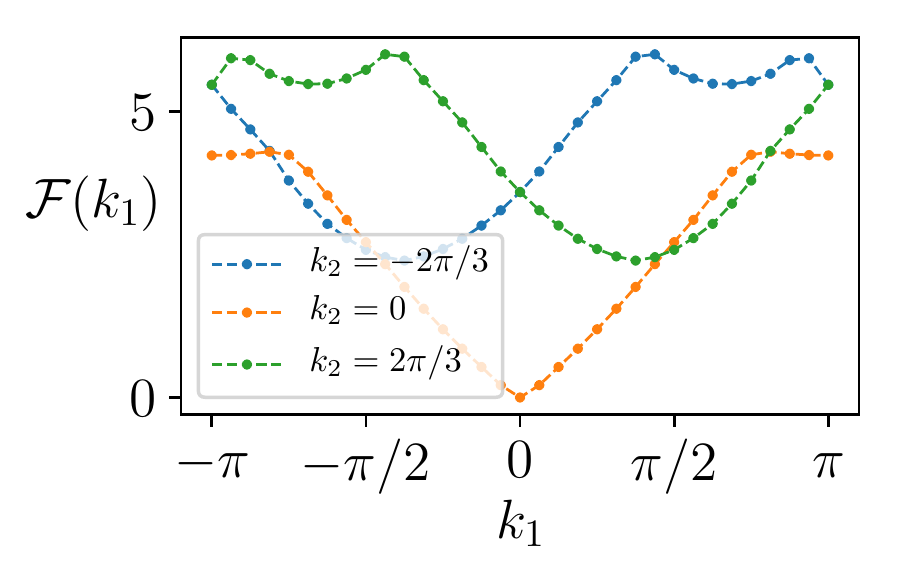} \put (0,62) {\footnotesize{(c)}} \end{overpic}
\caption{(a) Pseudospin gap as function of inverse system size for
finite cylinders of circumference $W=3$. Red dashed line is a
linear fit, while the green dashed line is a fit to $\Delta =
a/L$. We note that the value of the gap for the
largest system size of $L=40$ is less reliable, as the energy
extrapolation procedure is less accurate for this system size.
(b,c) Pseudospin structure factor obtained for a finite cylinder
with $W=3$ and length $L=32$ with respect to a site in the
middle of the system. (b) Non-interacting partons in the
mean-field band structure with $\Phi=\pi$; (c) DMRG.}
    \label{fig:Ny3}
\end{figure}

The finite-size behavior of the gap for $W=3$ is shown in
Fig.~\ref{fig:Ny3}(a). Although the results for the gap are not
fully conclusive, they are consistent with either a vanishing or a
very small gap. Here, a bond dimension of up to $M=8000$ was used,
resulting in a truncation error of $\epsilon_{\rm tr}\simeq
5\cdot10^{-5}$ for the ground state. Since the truncation errors
in the $S^z=1$ sector were slightly higher, to obtain a more
accurate value for the gap we performed an extrapolation of the
energy with truncation error in each sector before subtracting the
two (see Supplementary Material~\cite{SM} for further details).

To understand the nature of the state in this case, we consider
the static SU(4)-pseudospin structure factor,
\begin{equation}
    \mathcal{F}(\vec{k}) = \sum_{i} e^{i \vec{k}\cdot (\vec{r}_i-\vec{r}_{i_0})} \sum_\alpha \langle  \mcS^\alpha_{i} \cdot \mcS^\alpha_{i_0}  \rangle,
    \label{eq:StructureFactor}
\end{equation}
where $\vec{r}_i$, $\vec{r}_{i_0}$ denote the positions of the
sites $i$, $i_0$. For a gapless state with a parton Fermi surface,
the structure factor is expected to exhibit cusps at particular
momenta corresponding to the ``2$k_F$" values of the Fermi sea.
Fig.~\ref{fig:Ny3}(c) shows the structure factor calculated in the
ground state of a length $L=32$ cylinder using DMRG. Comparing it
to the structure factor calculated for the mean-field ansatz with
$\Phi=\pi$ using Wick's theorem (Fig.~\ref{fig:Ny3}(b)), we
observe good qualitative agreement and in particular see that the
cusps appear at the same momenta.

Finally, we note that starting from the mean-field ansatz, the
coupling to the gauge field may be numerically implemented by a
Gutzwiller projection, i.e. projecting the mean-field wavefunction
to a single-occupancy on each site. The correlations in the
resulting state can be probed using Monte Carlo sampling of the
projected wavefunction. Carrying out this projection, we find that
although we do not observe any symmetry breaking for $W=2,4$, the
power-law decay of the SU(4) pseudospin correlation function
$\sum_\alpha \langle  \mcS^\alpha_{i} \cdot
\mcS^\alpha_{i_0} \rangle$ in both cases agrees 
with the CFT
prediction. This result suggests that the Gutzwiller projection does
not capture the effect of the Umklapp interactions which are
particularly important to the cylinder geometries with $W=2,4$.
For $W=3$, we verify that the cusps in the structure factor remain
at the same position in the momentum space as for the mean-field
ansatz. Further details and numerical results are given in the
Supplementary Material~\cite{SM}.

\emph{Discussion-} For the quasi-1d geometries with $W=2,3,4$, the
DMRG results agree well with the analysis based on the parton
mean-field ansatz plus possible Umklapp interactions. We emphasize
that the symmetry-allowed Umklapp interactions considered are all
particular to certain geometries ($W=2,4$). They are not expected to
appear in the 2d limit as there is no Fermi-surface nesting in the
2d band structure (shown in Fig. \ref{fig:TriangularLattice_Band}
(b)) at filling $\nu=1/4$. In the 2d limit, the U(1) gauge flux
$\Phi$ also does not affect the parton Fermi surface. Therefore,
we expect that the parton Fermi surface obtained from the
mean-field ansatz Eq.~\eqref{eq:parton_meanfield} is stable in the
2d limit and provides a good candidate for the ground state of the
SU(4)-symmetric Kugel-Khomskii model Eq.~\eqref{eq:HeisenbergModel}
on the triangular lattice.

In real materials with spin and orbital degrees of freedom, one
can only expect an approximate SU(4) peseudospin symmetry. A small
SU(4)-symmetry-breaking perturbation is expected to split the
4-fold degeneracy of the 2d parton Fermi surface. A more
comprehensive investigation of the stability of the parton Fermi
surface to SU(4)-symmetry-breaking perturbations and other
none-Kugel-Khomskii-type interactions will be left for future
studies.

\acknowledgements

This research is funded in part by the Gordon and Betty Moore
Foundation through Grant GBMF8690 to UCSB to support the work of
A.K. 
C.X. is supported by NSF Grant No. DMR-1920434, the David and Lucile Packard Foundation, and the Simons Foundation.
Use was made of the computational facilities administered by
the Center for Scientific Computing at the CNSI and MRL (an NSF
MRSEC; DMR-1720256) and purchased through NSF CNS-1725797.

\bibliography{SU4}

\clearpage


\onecolumngrid
\setcounter{equation}{0}
\setcounter{figure}{0}
\renewcommand{\theequation}{S\arabic{equation}}
\renewcommand{\thefigure}{S\arabic{figure}}

\renewcommand{\thesection}{S\arabic{section}}
\renewcommand{\thesubsection}{\thesection.\arabic{subsection}}
\renewcommand{\thesubsubsection}{\thesubsection.\arabic{subsubsection}}

\begin{center}
{\Large\bfseries Supplementary Material}
\end{center}

\section{\boldmath Parton band structure for quasi-1d geometries with $W=2,3,4$ }
\begin{figure}[ht]
\begin{tabular}{c c c}
     \begin{overpic}[width=0.27\columnwidth]{BandsNy2_0Flux_LF2.pdf} \put (0,65) {\footnotesize{(a)}} \end{overpic} &
     \begin{overpic}[width=0.27\columnwidth]{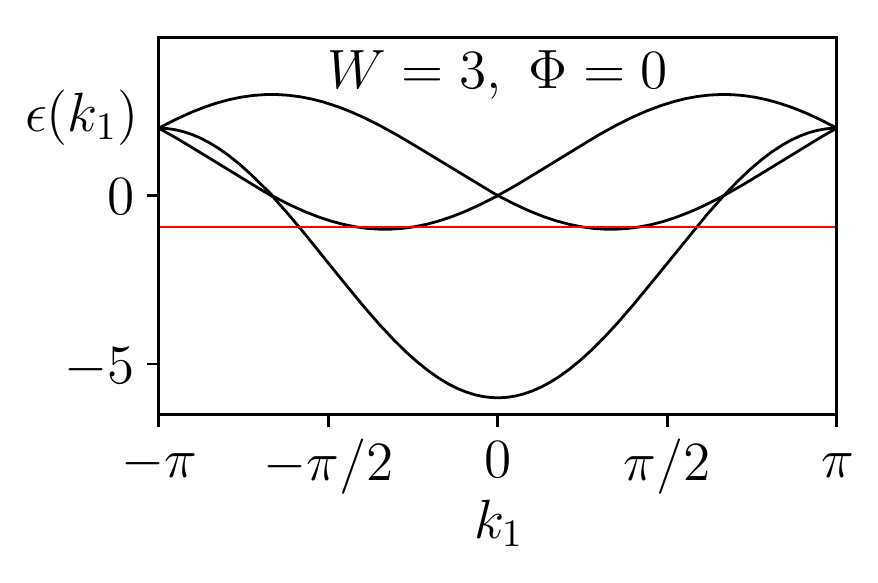} \put (0,65) {\footnotesize{(b)}} \end{overpic} &
     \begin{overpic}[width=0.27\columnwidth]{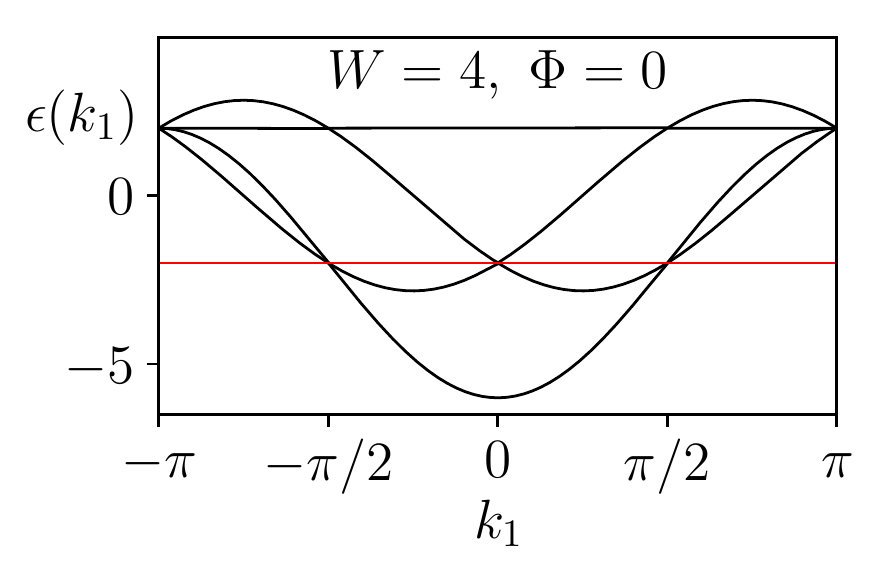} \put (0,65) {\footnotesize{(c)}} \end{overpic} \\
     \begin{overpic}[width=0.27\columnwidth]{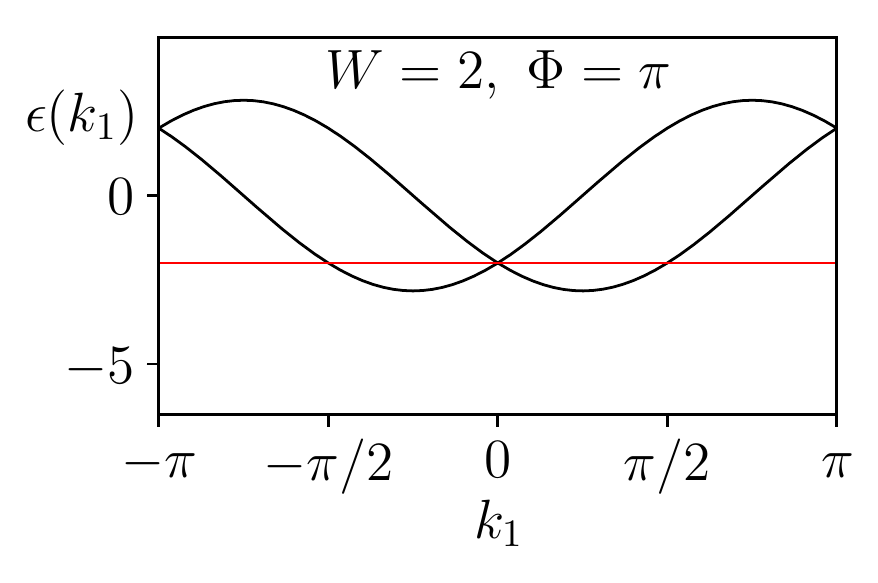} \put (0,65) {\footnotesize{(d)}} \end{overpic} &
     \begin{overpic}[width=0.27\columnwidth]{BandsNy3_PiFlux_LF2.pdf} \put (0,65) {\footnotesize{(e)}} \end{overpic} &
     \begin{overpic}[width=0.27\columnwidth]{BandsNy4_PiFlux_LF2.pdf} \put (0,65) {\footnotesize{(f)}} \end{overpic}
     \end{tabular}
\caption{Parton mean-field band structure for $W=2,3,4$ for
$\Phi=0$ (upper pannel) and $\Phi=\pi$ (lower pannel). The red
line indicates the Fermi level.}
    \label{fig:Quasi1DBands}
\end{figure}

In the main text, we focused on the 1d parton band structure with
$\Phi=0$ for $W=2$, and $\Phi=\pi$ for $W=3,4$. Here, for
completeness, we present the band structure for the complementary
choice of $\Phi$ for each $W$ (see Fig. \ref{fig:Quasi1DBands}).
Comparing the two possible scenarios for each $W$, we see that the
$\Phi$s discussed in the main text have less partially occupied
bands, and therefore, we expect them to be more stable. In
addition, we find that that the complimentary choices of $\Phi$
are not compatible with the DMRG study, even if symmetry-allowed
interactions are considered.

\subsection{Fermi momenta in the quasi-1d geometries}

The Fermi momenta can be calculated directly from the mean-field
Hamiltonian in Eq.~\eqref{eq:parton_meanfield} of the main text.
In all the cases we consider, these Fermi momenta are also
completely fixed by the symmetries $\mathcal{T}$, $C_2$ and
$\mathcal{M}$ which act on the two momenta $k_{1,2}$ as
\begin{equation}
   \mathcal{T}: k_{1,2} \rightarrow -k_{1,2},\quad \mathcal{C}_2: k_{1,2} \rightarrow -k_{1,2},\quad \mathcal{M} : k_1 \rightarrow -k_1 +k_2,\ k_2 \rightarrow k_2 .
\end{equation}
For $W=2$ with $\Phi=0$, the single partially occupied band has
$k_2=0$ (see Fig. \ref{fig:Quasi1DBands} (a)). The parton filling
constraint requires the two Fermi points to differ in their
$k_1$-values by $\pi$. Either symmetry $\mathcal{T}$,
$\mathcal{M}$, or $\mathcal{C}_2$ maps the two Fermi points into
each other and, therefore, fixes their values to be $\pm \pi/2$.

For $W=3,4$ with $\Phi=\pi$, the two bands that are (partially)
occupied by the partons have crystal momenta $k_2 = \pm \pi/W$
(see Fig. \ref{fig:Quasi1DBands} (e) and (f)). They transform into
each other under the time-reversal symmetry $\mathcal{T}$ or
$C_2$. The mirror symmetry $\mathcal{M}$ preserves $k_2$ and,
therefore, maps each partially occupied band to itself and
interchanges the two Fermi points in each band. Taking the parton
filling constraint into account, we conclude that the four Fermi
momenta in the case of $W=3,4$ are fixed to be at $\pm \pi/(2W)
\pm \pi W/8 $.

This symmetry-based analysis ensures that the values of the Fermi
momenta, in all the cases we focus on, are stable against small
deformations to the mean-field ansatz
Eq.~\eqref{eq:parton_meanfield}.

\section{\boldmath Umklapp interactions for the quasi-1d geometries with $W=2$ and $W=4$}
\subsection{\boldmath Quasi-1d geometry with $W=2$}
As explained in the main text, the parton mean-field ansatz
(including the coupling to U(1) gauge field) yields the $N_f=4$
QED$_2$ (or equivalently the SU$(4)_1$ CFT) description of the
quasi-1d geometry with $W=2$:
\begin{align}
    \mathcal{L}_{{\rm QED},W =2} = \sum_{m=1}^4 \left[ \psi^\dag_{L,m}(i\partial_0 - a_0 - i\partial_1 + a_1)\psi_{L,m}
    +
    \psi^\dag_{R,m}(i\partial_0 - a_0 + i\partial_1 - a_1)\psi_{R,m}
    \right].
\end{align}
The Umklapp interaction in Eq.~\eqref{eq:2leg_Interaction} of the
main text can be rewritten as
\begin{align}
   \mathcal{L}^{\rm int}_{W =2, \Phi=0}  = \left(\sum_{m} \psi^\dag_{L,m} \psi_{R,m}  \right)^2 + h.c = \frac{1}{2} \sum_{\alpha=1}^6 \left( \sum_{m_1,m_2=1}^4 \psi^\dag_{L,m_1}  M^{\alpha}_{m_1m_2} \psi^\dag_{L,m_2}  \right) \left( \sum_{m_3,m_4=1}^4\psi_{R,m_3}  M^{\alpha}_{m_3m_4} \psi_{R,m_4}  \right) +h.c.,
   \label{eq:2leg_Interaction_rewritten}
\end{align}
where $M^{\alpha}$ ($\alpha=1,2,...,6$) are $4\times 4$ matrices:
\begin{align}
    & M^1 = \left(\begin{array}{cccc}
        0 & -i & 0 & 0   \\
        i & 0  & 0 & 0   \\
        0 & 0  & 0 & 0   \\
        0 & 0  & 0 & 0
    \end{array}\right),~~~
    M^2 = \left(\begin{array}{cccc}
        0 & 0  & -i & 0   \\
        0 & 0  & 0 & 0   \\
        i & 0  & 0 & 0   \\
        0 & 0  & 0 & 0
    \end{array}\right),~~~
    M^3 = \left(\begin{array}{cccc}
        0 & 0  & 0 & -i   \\
        0 & 0  & 0 & 0   \\
        0 & 0  & 0 & 0   \\
        i & 0  & 0 & 0
    \end{array}\right),~~~ \nonumber \\
    & M^4 = \left(\begin{array}{cccc}
        0 & 0  & 0 & 0   \\
        0 & 0  & -i & 0   \\
        0 & i  & 0 & 0   \\
        0 & 0  & 0 & 0
    \end{array}\right),~~~
    M^5 = \left(\begin{array}{cccc}
        0 & 0  & 0 & 0   \\
        0 & 0  & 0 & -i   \\
        0 & 0  & 0 & 0   \\
        0 & i  & 0 & 0
    \end{array}\right),~~~
    M^6 = \left(\begin{array}{cccc}
        0 & 0  & 0 & 0   \\
        0 & 0  & 0 & 0   \\
        0 & 0  & 0 & -i   \\
        0 & 0  & i & 0
    \end{array}\right).
\end{align}
Under the SU(4) pesudospin rotation, the 6 matrices $M^\alpha$
transform as the 6-dimensional representation of SU(4) (i.e. the
vector representation of SO(6)). The $N_f=4$ QED$_2$ can be viewed
as a realization of the CFT coset construction SU(4)$_1 =
$U(4)$_1$/U(1)$_4$ where the U(4)$_1$ simply corresponds to a
4-component massless Dirac fermion without coupling to the U(1)
gauge field. From this perspective, the field $\left(
\sum_{m,m'=1}^4 \psi^\dag_{L,m}  M^{\alpha}_{mm'} \psi^\dag_{L,b}
\right)$ can be viewed as creating a primary field carrying the
6-dimensional representation of SU(4) in the left-moving sector of
the SU(4)$_1$ CFT while its gauge charge under U$(1)$ is
``quotient" out by U(1)$_4$ in the coset-construction language. A
similar reasoning applies to the field $\left( \sum_{m,m'=1}^4
\psi^\dag_{R,m}  M^{\alpha}_{mm'} \psi^\dag_{R,b}  \right)$.
Therefore, as mentioned in the main text, we can interpret the
Umklapp interaction Eq.~\eqref{eq:2leg_Interaction} as the
back-scattering between left-moving and right-moving primary
fields in the SU(4)$_1$ CFT, both carrying the 6-dimensional
representation of SU(4).

As we explained in the main text, the Umklapp interaction has
scaling dimension 1 and, hence, is a relevant perturbation in the
SU(4)$_1$ CFT.  As it runs strong under the renormalization-group
flow, the system can enter a phase where the field
$\left(\sum_{m=1}^4 \psi^\dag_{L,m} \psi_{R,m}\right)$ condenses.
In this condensate, the Umklapp interaction effectively generates
the following mass term for the Dirac fermions:
\begin{align}
    \mathcal{L}_{{\rm mass}, W= 2} = \phi \sum_{m=1}^4 \psi^\dag_{L,m} \psi_{R,m} + \phi^* \sum_{m=1}^4 \psi^\dag_{R,m} \psi_{L,m},
\end{align}
where the complex number $\phi$ is essentially the (non-zero)
vacuum expectation $\langle\sum_{m=1}^4 \psi^\dag_{L,m}
\psi_{R,m}\rangle$. This mass term gaps out all the low-energy
excitations in the SU(4)$_1$ and spontaneously breaks the
$T_1$-translation symmetry with a doubled unit cell. This mass
term still respects the SU(4)-pseudospin-rotation symmetry and the
TR symmetry $\mathcal{T}$. When $\phi$ is not purely real, the
mirror symmetry $\mathcal{M}$ and the rotation symmetry $C_2$ are
also spontaneously broken. Indeed, the DMRG results for $W=2$
(shown in Fig. \ref{fig:NyEven} (b) of the main text) do exhibit,
in addition to the spontaneous breaking of $T_1$, the breaking of
both symmetries $\mathcal{M}$ and $C_2$.

\subsection{\boldmath Quasi-1d geometry with $W=4$}
For the quasi-1d geometry with $W=4$, the parton mean-field ansatz
(including the coupling to U(1) gauge field) yields the $N_f=8$
QED$_2$ (or equivalently the SU$(8)_1$ CFT):
\begin{align}
    \mathcal{L}_{{\rm QED},W =4} = \sum_{m=1}^4 \sum_{n=1}^2 \left[ \psi^\dag_{L,n,m}(i\partial_0 - a_0 - i\partial_1 + a_1)\psi_{L,n,m}
    +
    \psi^\dag_{R,n,m}(i\partial_0 - a_0 + i\partial_1 - a_1)\psi_{R,n,m}
    \right],
\end{align}
where the 4-fold pseudospin index $m$ and the 2-fold band index
$n$ together form the index for the fundamental representation of
SU(8) in the SU$(8)_1$ CFT. The Umklapp interactions
\begin{align}
\left(\sum_{m=1}^4 \psi^\dag_{L,n,m} \psi_{R,n,m}  \right) \left(\sum_{m'=1}^4 \psi^\dag_{L,n',m'} \psi_{R,n',m'}  \right)+h.c.,
\end{align}
with $n,n'=1,2$ preserve all the symmetries of the model but not
necessarily the full SU(8) enhanced symmetry of the SU$(8)_1$ CFT.
Nevertheless, we can use the knowledge of the SU$(8)_1$ CFT to
analyze the scaling dimension of the Umklapp interactions. Via a
similar rewriting as Eq.~\eqref{eq:2leg_Interaction_rewritten} and
via the coset construction SU(8)$_1 = $U(8)$_1$/U(1)$_8$, we can
identify the Umklapp terms as the back-scattering between
left-moving and right-moving primary fields in the SU(8)$_1$ CFT,
both carrying the 28-dimensional representation of SU(8).

As explained in the main text, all such Umklapp interactions have
the scaling dimension $3/2$ and hence are relevant under the
renormalization group. At low energy, the Umklapp terms can drive
the system into a phase with non-zero vacuum expectation values of
both fields $\left(\sum_{m=1}^4 \psi^\dag_{L,1,m} \psi_{R,1,m}
\right)$ and $\left(\sum_{m=1}^4 \psi^\dag_{L,2,m} \psi_{R,2,m}
\right)$. In this phase, the Umklapp interactions effectively
generate the following mass terms for the Dirac fermions:
\begin{align}
    \mathcal{L}_{{\rm mass}, W= 4} =  \phi_1 \sum_{m=1}^4 \psi^\dag_{L,1,m} \psi_{R,1,m} + \phi_1^* \sum_{m=1}^4 \psi^\dag_{R,1,m} \psi_{L,1,m}
    +
     \phi_2 \sum_{m=1}^4 \psi^\dag_{L,2,m} \psi_{R,2,m} + \phi_2^* \sum_{m=1}^4 \psi^\dag_{R,2,m} \psi_{L,2,m} ,
\end{align}
where the complex numbers $\phi_{1,2}$ are essentially the vacuum
expectation values $\langle\sum_{m=1}^4 \psi^\dag_{L,1,m}
\psi_{R,1,m}  \rangle$ and $\langle \sum_{m=1}^4 \psi^\dag_{L,2,m}
\psi_{R,2,m} \rangle$ respectively. These mass terms gap out all
the low-energy excitations in the SU(8)$_1$ CFT and spontaneously
breaks the $T_1$-translation symmetry with a doubled unit cell.
They still respect the SU(4)-pseudospin-rotation symmetry. The TR
symmetry $\mathcal{T}$ requires $\phi_1 = \phi_2$. If either
$\phi_1$ or $\phi_2$ is not purely real, the mirror symmetry
$\mathcal{M}$ and rotation symmetry $C_2$ are spontaneously broken
as well. Indeed, the DMRG results for $W=4$ (shown in Fig.
\ref{fig:NyEven} (d) of the main text) do exhibit, in addition to
the spontaneous breaking of $T_1$, the breaking of both symmetries
$\mathcal{M}$ and $C_2$.

\section{Gutzwiller projected wavefunctions}

Starting from the mean-field ansatz, the single-occupancy
constraint on each site can be implemented by means of a
Gutzwiller projection. Correlations in the resulting state can be
probed using Monte Carlo sampling of the projected wavefunction.
In the following, we will show that for $W=3$, the correlations
obtained from the Gutzwiller-projected wavefunction are in good
agreement with the correlations of the exact ground state, as
obtained from DMRG. For $W=2$ and $W=4$, we find that the
Gutzwiller state does not break translational symmetry. However,
it exhibits dominant correlations consistent with the CFT
prediction.

\begin{figure}
    \begin{overpic}[width=0.29\columnwidth]{SpinCorrelations_FF_Pi_L32_LF_F.pdf} \put (0,62) {\footnotesize{(a)}} \end{overpic} \qquad\qquad
    \begin{overpic}[width=0.29\columnwidth]{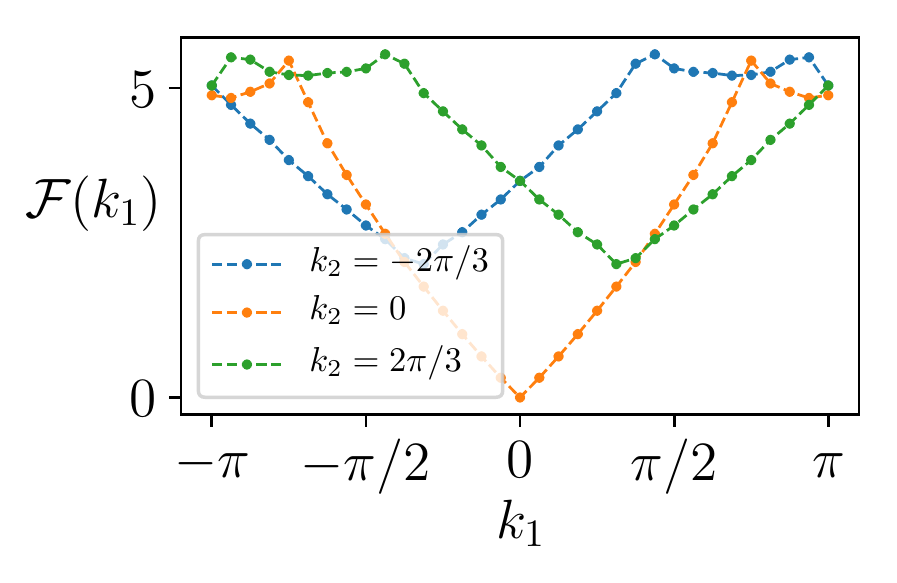} \put (0,62) {\footnotesize{(b)}} \end{overpic}
\caption{Pseudospin structure factor obtained for a finite
cylinder with $W=3$ and length $L=32$ with respect to a site
in the middle of the system. (a) Non-interacting parton Fermi sea
with $\Phi=\pi$; (b) Gutzwiller projection.}
    \label{fig:Ny3Correlations}
\end{figure}
For $W=3$, we calculate the static structure factor (see
Eq.~\eqref{eq:StructureFactor} in the main text) in the
Gutzwiller-projected state and verify that the cusps in the
structure factor remain at the same values of the momenta as for
the mean-field ansatz. This can be seen in
Fig.~\ref{fig:Ny3Correlations}.

\begin{figure}[ht]
    \begin{overpic}[width=0.35\columnwidth]{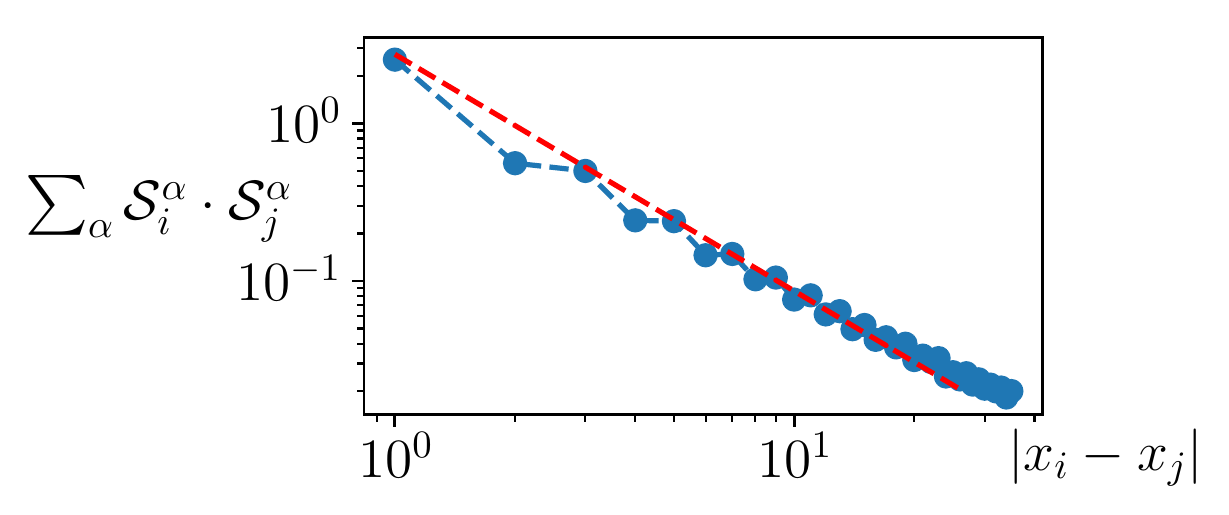} \put (0,40) {\footnotesize{(a)}} \end{overpic} \qquad
    \begin{overpic}[width=0.35\columnwidth]{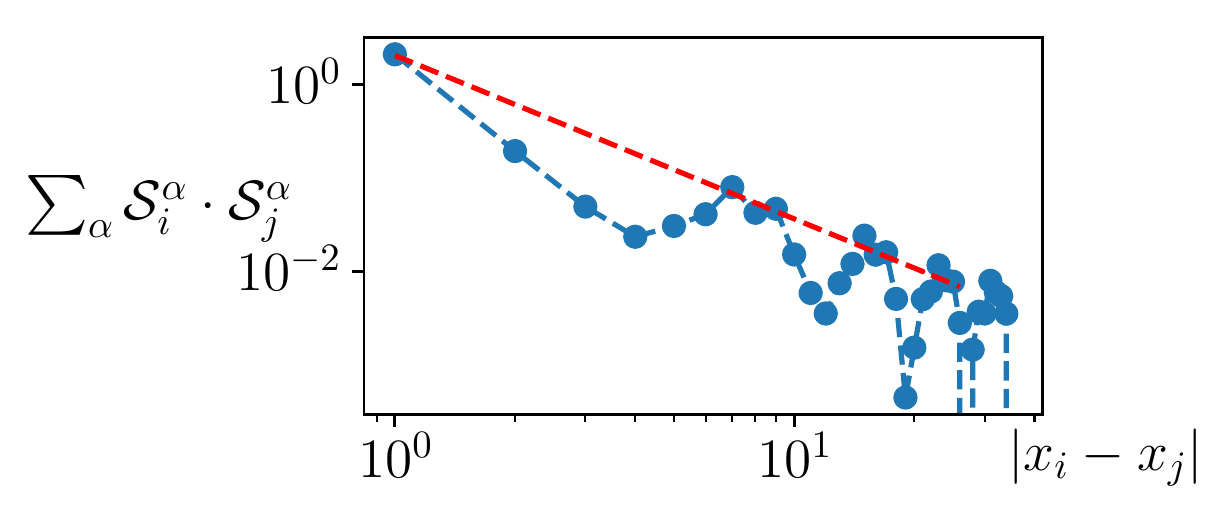} \put (0,40) {\footnotesize{(b)}} \end{overpic}
\caption{Real space pseudospin correlations for the Gutzwiller
projected wavefuntions in a system with periodic boundary
conditions, length $L=110$ and circumference $W=2$ in (a) and
$W=4$ in (b). Red dashed lines corresponds to the exponents
expected from the CFT, i.e. $3/2$ in (a) and $7/4$ in (b).}
    \label{fig:GW_Correlations_EvenW}
\end{figure}

Turning now to $W=2$ and $W=4$, we observe a power-law decay of
the correlations for both cases, thus suggesting that the system
is described by a gapless field theory. In
Fig.~\ref{fig:GW_Correlations_EvenW}, we show that the exponents
of the decay are consistent with $3/2$ for the case of $W=2$ and
$7/4$ for $W=4$. As we explain below, these values are consistent
with the field theory predictions discussed in the main manuscript
in the absence of the relevant Umklapp interactions, which
apparently are not captured on the level of Gutzwiller projection.
Specifically, the two field theories are SU(4)$_1$ CFT for $W=2$
and SU(8)$_1$ CFT for $W=4$ in the absence of Umklapp
interactions.

For the $W=2$ case, we consider gapless $N_f=4$ QED$_2$ or the
SU(4)$_1$ CFT without Umklapp interactions. The parton
decomposition Eq.~\ref{eq:parton_decomposition} suggests that the
SU(4) pseudospin correlation function $\sum_\alpha \langle
\mcS^\alpha_{i} \cdot \mcS^\alpha_{j}  \rangle$ should share the
same scaling exponent with the correlations function $\langle
\sum_m \psi_{L,m} (x) \psi_{L,m}^\dag (0) \rangle \langle
\sum_{m'} \psi_{R,m'}^\dag (x) \psi_{R,m'} (0) \rangle $ in the
SU(4)$_1$ CFT. $\langle \sum_m \psi_{L,m} (x) \psi_{L,m}^\dag (0)
\rangle$ corresponds to the correlator of the left-moving primary
field carrying the fundamental representation of SU(4) which
scales as $\sim |x|^{-\frac{3}{4}}$. Similarly, its right-moving
counterpart $\langle \sum_{m'} \psi_{R,m'}^\dag (x) \psi_{R,m'}
(0) \rangle $ also scales as $\sim |x|^{-\frac{3}{4}}$. Therefore,
based on the SU(4)$_1$ CFT, we expect the scaling $\sum_\alpha
\langle  \mcS^\alpha_{i} \cdot \mcS^\alpha_{j}  \rangle \sim |x_i
- x_j|^{-\frac{3}{2}} $ which is in agreement with the result for the
Gutzwiller projected state shown in
Fig.~\ref{fig:GW_Correlations_EvenW}(a). Here, $|x_i - x_j|$
denote the distance between site $i$ and $j$ in the $T_1$
direction. In the parton band structure with $\Phi=0$ for $W=2$, the
two Fermi points differ in their $k_1$-values by $\pi$. We
therefore expect a ``2-site-periodic" modulation on top of
the power-law decay in the SU(4)-pseudospin correlation function,
which is indeed observed in Fig.~\ref{fig:GW_Correlations_EvenW}(a).

For the $W=4$ case, we expect that the state is described by the
gapless $N_f=8$ QED$_2$ or the SU(8)$_1$ CFT without  Umklapp
interactions. The SU(4)-pseudospin correlation function
$\sum_\alpha \langle  \mcS^\alpha_{i} \cdot \mcS^\alpha_{j}
\rangle$ should now share the same scaling exponent with the
correlations function $\langle \sum_m \psi_{L,m} (x)
\psi_{L,m}^\dag (0) \rangle \langle \sum_{m'} \psi_{R,m'}^\dag (x)
\psi_{R,m'} (0) \rangle $ in the SU(8)$_1$ CFT. $\langle \sum_m
\psi_{L,m} (x) \psi_{L,m}^\dag (0) \rangle$ corresponds to the
correlator of the left-moving primary field carrying the
fundamental representation of SU(8) which scales as $\sim
|x|^{-\frac{7}{8}}$. Similarly, its right-moving counterpart
$\langle \sum_{m'} \psi_{R,m'}^\dag (x) \psi_{R,m'} (0) \rangle $
also scales as $\sim |x|^{-\frac{7}{8}}$. Therefore, based on the
SU(8)$_1$ CFT, we expect the scaling $\sum_\alpha \langle
\mcS^\alpha_{i} \cdot \mcS^\alpha_{j}  \rangle \sim |x_i -
x_j|^{-\frac{7}{4}} $ which is in agreement with the result for the
Gutzwiller projected state shown in
Fig.~\ref{fig:GW_Correlations_EvenW}(b). In the parton band
structure with $\Phi=\pi$ for $W=4$, the pairwise differences of
the four Fermi points in their $k_1$-values are commensurate to an
8-site unit cell along $T_1$ direction. We therefore expect
an ``8-site-periodic" modulation on top of the power-law decay in
the SU(4)-pseudospin correlation function, which is indeed observed
in Fig.~\ref{fig:GW_Correlations_EvenW}(b).

\section{Additional DMRG Results}

\subsection{\boldmath Energy extrapolation with truncation error for $W=3$}

\begin{figure}[ht]
    \begin{overpic}[width=0.33\columnwidth]{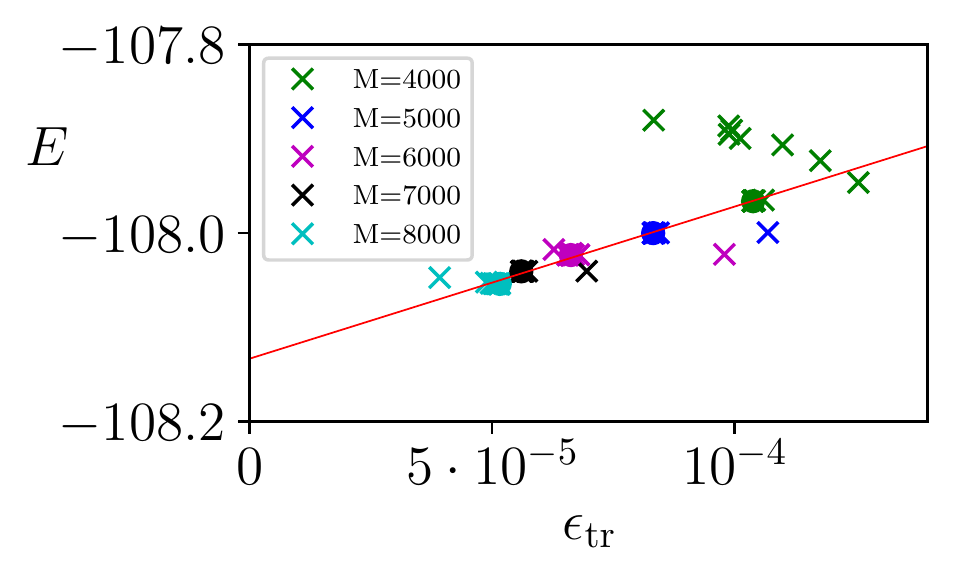} \put (-5,55) {\footnotesize{(a)}} \end{overpic} \qquad
    \begin{overpic}[width=0.33\columnwidth]{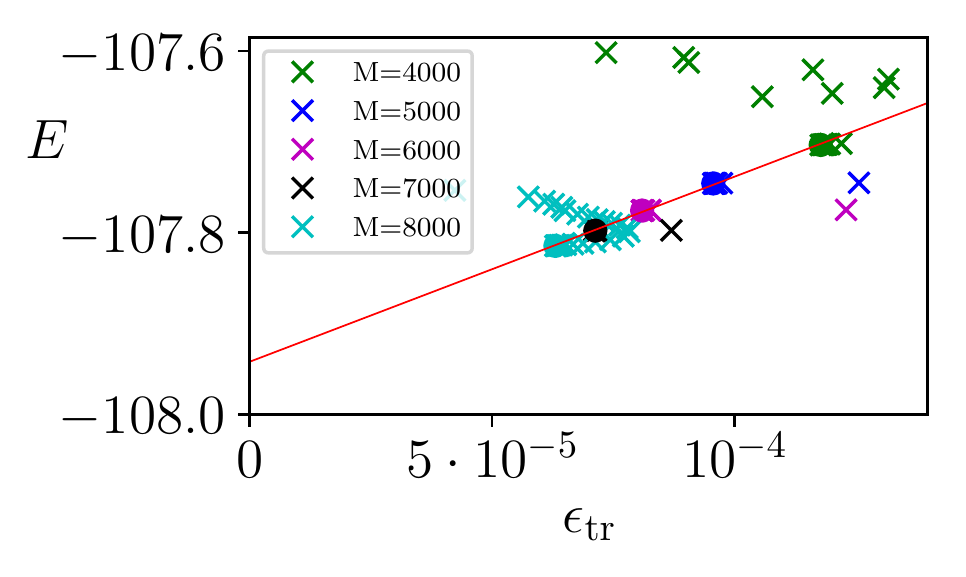} \put (-5,55) {\footnotesize{(b)}} \end{overpic}
\caption{Energy extrapolation with truncated weight for a cylinder
of circumference $W=3$ and length $L=32$ (a) in the $\left(S^z,
V^z, S^z V^z\right)=\left(0,0,0\right)$ sector and (b) in the
$\left(S^z, V^z, S^z V^z\right)=\left(1,0,-1/2\right)$ sector.
Each cross on the plot corresponds to the energy, $E$, and
truncated weight, $\epsilon_{\rm tr},$ for a given DMRG sweep,
with the color of the point indicating the the maximal bond
dimension $M$ for that sweep. The last sweep at each bond
dimension is indicated by a square of the same color. These points
are the ones used for the linear extrapolation of the energy down
to zero truncated weight (the red line). }
    \label{fig:EnergyExtrap}
\end{figure}

To obtain a more accurate estimate for the gap for cylinders of
circumference $W=3$, we first extrapolate the energy in each
quantum-numbers sector down to zero truncated weight. The gap is
then calculated by subtracting the extrapolated energies. In
Fig.~\ref{fig:EnergyExtrap} we show an example of the
extrapolation procedure for a cylinder of length $L=32$.

\subsection{\boldmath Breaking of translation invariance for $W=4$}

\begin{figure}[ht]
    \begin{overpic}[width=0.45\columnwidth]{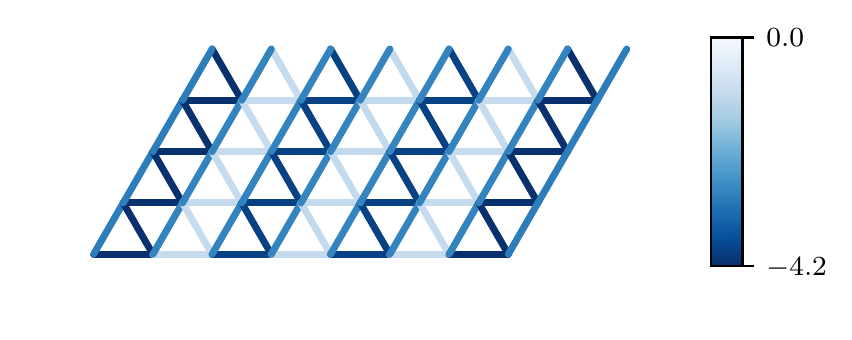} \put (0,35) {\footnotesize{(a)}} \end{overpic} \qquad
    \begin{overpic}[width=0.45\columnwidth]{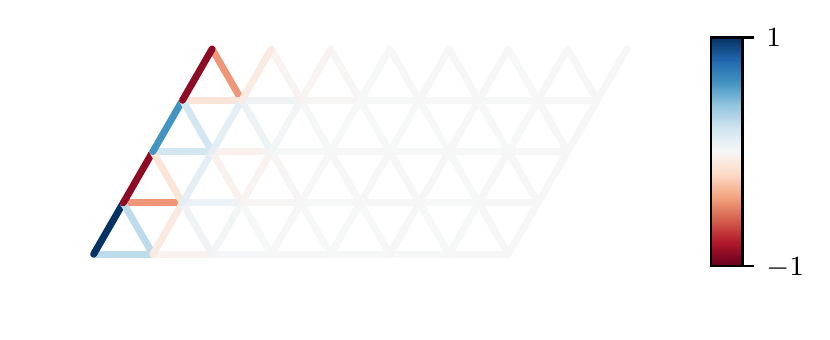} \put (0,35) {\footnotesize{(b)}} \end{overpic}
\caption{(a) Expectation values of $\sum_\alpha \mcS^\alpha_i\cdot
\mcS^\alpha_j $ for a cylinder of circumference $W=4$ and length
$L=8$. (b) Difference in the bond expectation values upon doubling
the strength of the coupling $J$ on a vertical bond on the edge of
the cylinder.}
    \label{fig:Ny4BondsDiff}
\end{figure}

To rule out the possibility that the ground state obtained in DMRG
is a superposition of two states with broken translation symmetry
along the circumference (i.e. along $T_2$), we double the strength
of the coupling on one of the vertical bonds on the first rung of
the cylinder. We observe only a relatively small change in the
expectation values of the bonds close to the perturbed bond, and a
rapid decay of the difference away from it, as can be seen in
Fig.~\ref{fig:Ny4BondsDiff}.

\end{document}